\documentclass[9pt,twocolumn,twoside]{opticajnl}
\journal{opticajournal} 

\usepackage{graphicx}
\usepackage{dcolumn}
\usepackage{bm}
\usepackage{hhline}
\usepackage{array}
\usepackage{xcolor}
\usepackage{soul}
\usepackage{wrapfig}
\usepackage{float}

\begin{document}



\title{Integrated Mode-Hop-Free Tunable Lasers at 780 nm for Chip-Scale Classical and Quantum Photonic Applications}

\author[1,$\dag$]{Joshua E. Castro}
\author[2,$\dag$]{Eber Nolasco-Martinez}
\author[1,3,$\dag$]{Paolo Pintus}
\author[4]{Zeyu Zhang}
\author[4]{Boqiang Shen}
\author[1]{Theodore Morin}
\author[1]{Lillian Thiel}
\author[5]{Trevor J. Steiner}
\author[1]{Nicholas Lewis}
\author[1]{Sahil D. Patel}
\author[1,5]{John E. Bowers}
\author[2]{David M. Weld}
\author[1,*]{Galan Moody}

\affil[1]{Electrical and Computer Engineering Department, University of California, Santa Barbara, CA 93106, USA}
\affil[2]{Physics Department, University of California, Santa Barbara, CA 93106, USA}
\affil[3]{Physics Department, University of Cagliari, Monserrato, IT 09042, Italy}
\affil[4]{Nexus Photonics Inc., Goleta, CA 93117, USA}
\affil[5]{Materials Department, University of California, Santa Barbara, CA 93106, USA}

\affil[$\dag$]{These authors contributed equally to this work.}
\affil[*]{moody@ucsb.edu}


\date{\today}
\begin{abstract}
In the last decade, remarkable advances in integrated photonic technologies have enabled table-top experiments and instrumentation to be scaled down to compact chips with significant reduction in size, weight, power consumption, and cost. Here, we demonstrate an integrated continuously tunable laser in a heterogeneous gallium arsenide-on-silicon nitride (GaAs-on-SiN) platform that emits in the far-red radiation spectrum near $780$~nm, with $20$ nm tuning range, $< 6$ kHz intrinsic linewidth, and a $>40$ dB side-mode suppression ratio. The GaAs optical gain regions are heterogeneously integrated with low-loss SiN waveguides. The narrow linewidth lasing is achieved with an extended cavity consisting of a resonator-based Vernier mirror and a phase shifter. Utilizing synchronous tuning of the integrated heaters, we show mode-hop-free wavelength tuning over a range larger than $100$ GHz ($200$ pm). To demonstrate the potential of the device, we investigate two illustrative applications: (i)~the linear characterization of a silicon nitride microresonator designed for entangled-photon pair generation, and (ii)~the absorption spectroscopy and locking to the $D_1$ and $D_2$ transition lines of $^{87}$Rb. The performance of the proposed integrated laser holds promise for a broader spectrum of both classical and quantum applications in the visible range, encompassing communication, control, sensing, and computing.
\end{abstract}

 
\maketitle
\section{\label{sec:intro}Introduction}
\vspace{-5pt}

\begin{figure}[t]
\centering
\includegraphics[width=1\linewidth]{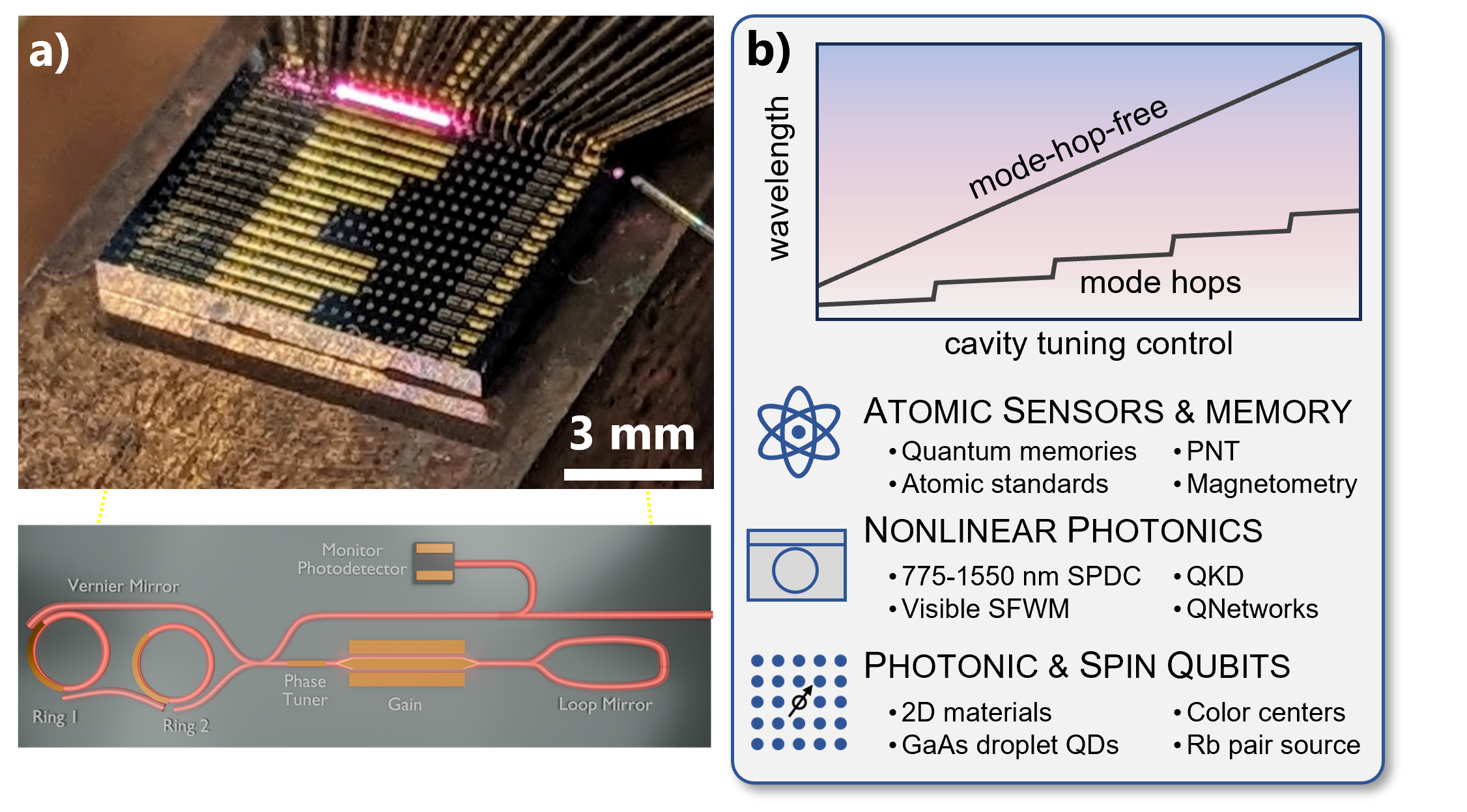}
\caption{\footnotesize \textbf{Mode-hop-free tunable semiconductor chip-scale lasers.} (a) Image of 15 tunable III-V lasers integrated on a single $\sim 25$ cm$^2$ chip. Depending on the specific laser channel, 765-795 nm light is coupled off-chip using a single-mode polarization-maintaining fiber. Each laser comprises a gain region, a tunable Vernier mirror, a phase-tuning element, and a photo-diode monitor, as shown in the bottom panel. (b) An algorithm to tune the Vernier mirror reflectivity and phase element enables $>100$ GHz mode-hop free-tuning, which has applications for atomic sensors and memories, nonlinear photonics, and quantum information.}
\vspace{-10pt}
\label{fig:intro}
\end{figure}

\begin{figure*}[ht!] 
\centering
  \includegraphics[width=0.99\linewidth]{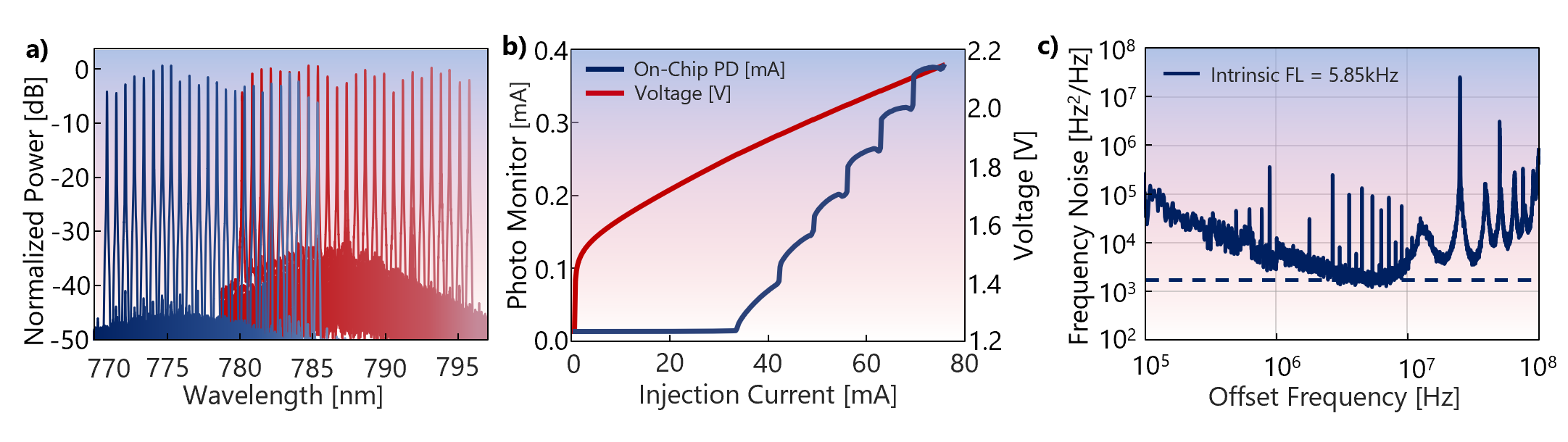} 
  \caption{\footnotesize \label{fig:laser} \textbf{Characterization of III-V laser performance.} (a) Tuning curves of two different laser cavities with III-V quantum well stacks of different thicknesses covering 765-795 nm laser operation 35-45 dB side-mode suppression. (b) Measurement of the laser output using the on-chip photodiode monitor (left axis) and drive voltage (right axis) versus gain injection current, demonstrating a laser threshold around 30 mA. (c) Representative frequency noise measurement demonstrating 5.85 kHz fundamental linewidth without additional stabilization or locking.}    
\end{figure*}

As the complexity and scale of quantum systems continue to grow, the role of integrated photonics has become increasingly pivotal~\cite{giordani2023integrated,pelucchi2022potential}. Recent innovations in hybrid and heterogeneous photonic platforms~\cite{tran2022extending,kaur2021hybrid,liang2021recent} have enabled the seamless integration of classical components---coherent light sources, modulators, and detectors---with quantum elements including quantum frequency combs, trapped ions, neutral atoms, solid-state quantum emitters, and quantum transducers~\cite{Moody_2022}. This fusion of classical and quantum integrated photonics not only accelerates near-term applications in quantum communications, sensing, and networking, but it also addresses many of the engineering challenges that the bulk optical systems are associated with, such as multiplexing, phase noise, system stability, and reliability~\cite{labonte2024integrated}. 

Narrow-linewidth tunable lasers---ubiquitous in classical and quantum applications---have massively benefited from integrated photonics in recent years. Traditionally, the external-cavity diode laser (ECDL) has been the most effective design to achieve $\sim$kilohertz linewidth single-mode output, wide wavelength tunability, and high power delivery; however, their large footprint, mechanical wavelength tuning mechanisms, and sensitivity to vibration and shock often restrict their usage to the confines of well-controlled lab settings. For technologies and applications that require low size, weight, and power (SWaP), recent demonstrations of hybrid and heterogeneous integrated III-V semiconductor lasers with silicon and silicon nitride photonics have dramatically expanded the possibilities for low SWaP coherent light delivery for classical and quantum applications~\cite{porcel2019silicon,tran2019tutorial,de2020heterogeneous}. These advances were initially driven by the need for high-bandwidth and low-cost interconnects for data communications at telecom wavelengths~\cite{pintus2023demonstration,xiang2021perspective,porter2023hybrid}, but they are now transforming many additional technology sectors including quantum information, time-keeping and remote sensing, and artificial intelligence~\cite{mahmudlu2023fully,zhou2023prospects,shastri2021photonics}.  

Expanding the useful wavelength range to the near-ultraviolet and visible spectrum would open additional avenues for applications requiring low SWaP. These applications include but are not limited to cooling, repumping, and control of atoms and molecules with complex spectra, nonlinear classical and quantum photonics, and solid-state quantum system control and readout~\cite{lai2020780,wang2023photonic}. Key performance metrics that need to be optimized for these applications are: (1) wide coarse tunability with fine mode-hop-free (MHF) tuning to map out optical atomic and molecular spectra, (2) sub-megahertz linewidths to address particular resonances and electronic transitions, (3) milliwatt-level output power across the tuning range for coherently driving nonlinear processes and to cool, trap, and manipulate atoms and molecules, and (4) single-mode operation with large side-mode suppression to minimize noise and parasitic processes. Significant progress has been made in recent years with advances in semiconductor growth and fabrication techniques and photonic integration methods, including kilohertz to megahertz linewidths in the visible spectrum, ranging from 405~nm up to 684~nm, using hybrid integration of semiconductor laser chips or optical amplifiers with bulk optics or integrated photonic cavities and switching networks~\cite{donvalkar2018self,savchenkov2020application,mashayekh2021silicon,franken2021hybrid,winkler2023silicon,winkler2024widely,chauhan2021visible}. Corato-Zanarella \textit{et al.} recently extended the operating wavelength to $785$~nm through hybrid integration of a Fabry-Perot semiconductor laser chip butt-coupled to a low-loss silicon nitride resonators for self-injection locking to achieve sub-10 kHz instantaneous linewidth with 12.5 nm coarse tuning, 34 GHz MHF tuning, side-mode suppression > 37 dB, and 10 mW output power~\cite{corato2023widely}.

In this work, we demonstrate a fully integrated, heterogeneous GaAs-on-SiN platform that enables mode-hop free continuous tuning over 100 GHz (200 pm) with $>15$ nm coarse tuning. Building on the previous success of Zhang \textit{et al.} in demonstrating coarse wavelength sweeps from 765~nm and 795~nm ~\cite{zhang2023photonic}, we implement a MHF tuning algorithm that simultaneously controls a dual-ring Vernier mirror and intra-cavity phase shifter~\cite{pintus2023demonstration}, enabling $\geq 100$ GHz mode-hop-free tuning. As shown in Fig.~\ref{fig:intro}, the integrated laser chip comprises 15 independently tunable lasers with GaAs quantum well (QW) gain sections bonded onto silicon nitride waveguides. We measure intrinsic linewidths $<6$ kHz, side-mode suppression ratio (SMSR) $>40$ dB, and over 10 mW output power. We demonstrate the capabilities of these lasers on two key applications. First we carry out linear spectroscopy of a silicon nitride microresonator designed for visible entangled-photon pair generation via spontaneous four-wave mixing at 780 nm. The tuning algorithm can adeptly map the $20$ GHz spaced modes with quality factor Q $\sim2$ million. In a second application, we perform Doppler-free spectroscopy of the D$_1$ and D$_2$ transitions of an $^{87}$Rb vapor. The exquisite tuning sensitivity easily identifies sub-Doppler hyperfine features, and the laser stability enables it to be locked to an absolute atomic frequency reference over the entire 10 minute measurement duration. We conclude with a discussion on future capabilities enabled by this work in atomic sensing and quantum memories, quantum control, nonlinear quantum photonics, and photonic and spin qubits.

\section{Device Layout and Laser Characterization}

The integrated visible laser, shown in optical image and design schematic in Fig.~\ref{fig:intro}, utilizes a GaAs multi-quantum well (QW) optical gain region heterogeneously integrated with SiN waveguides. This combination leverages the high gain and efficient light emitting properties of GaAs with the robust, CMOS compatible, low loss SiN material platform, thus enabling the realization of compact, high performance fully integrated lasers~\cite{lin2022monolithically}. The lasers demonstrated in this work are designed and fabricated by Nexus Photonics, Inc. in their heterogeneous GaAs-on-SiN device platform. 



The tunable laser utilizes an external cavity design based on low-loss SiN, which is crucial in achieving narrow-linewidth and stable single-mode operation~\cite{tran2019tutorial,van2020ring}. The back-end of the laser cavity consists of an external cavity Sagnac loop mirror within which there are two tandem add-drop ring resonators with slightly different radii. This design enables a large Vernier free spectral range (FSR), calculated as |$FSR_1$|*|$FSR_2$|/|$FSR_1 - FSR_2$|, where the subscripts correspond to ring 1 and 2, respectively~\cite{tran2019ring}. An FSR larger than the bandwidth of the gain medium ensures single longitudinal mode lasing and improves the SMSR. The Vernier mirror design also increases the photon lifetime inside the laser cavity thanks to the high quality factor of the SiN ring resonators, resulting in significantly reduced laser linewidths compared to other single-mode laser designs while still maintaining a small on-chip footprint~\cite{tran2019tutorial}. By thermally controlling the Vernier mirror and the intra-cavity phase shifter, the laser wavelength can be effectively controlled and shifted by more than 15 nm. The metal heater layer is deposited above the oxide cladding of the SiN waveguides. The front mirror consists of a Sagnac-loop mirror with a reflectivity of $100~\%$ over a broad wavelength range ($>$15 nm). Light is coupled out of the cavity via a multi-mode interferometer, where $15\%$ of light is directed to an integrated photodiode (PD), composed of the same III-V structure as the gain region, but operated in reverse bias for on-chip power monitoring. The remaining light not directed into the PD exits the chip via an angled edge coupler with optimized mode overlap with a lensed single-mode fiber.

All characterization and demonstrations were performed with the laser chip on a temperature controlled stage set to $25$ \textdegree C. Electrical contacts to the gain, PD, and thermo-optic tuning elements were made using a multi-point probe card shown in Fig.~\ref{fig:intro}. Light generated from the laser was coupled off-chip into a polarization-maintaining lensed optical fiber with a spot size of $2$ $\mu$m and a working distance of $10$ $\mu$m. The fiber output was then sent to a fiber-coupled optical isolator with over $35$ dB of isolation to minimize back-reflections into the laser. Because we utilized fiber coupling, as opposed to packaged devices, $10\%$ of the power after the isolator was routed to an active auto-alignment fiber controller (Thorlabs KNA-VIS Nanotrak) to maximize the power collected by the lensed fiber and to reduce spurious losses from the facet to fiber coupling. The transmitted power is then routed to a 3-dB fiber splitter, whereby one output is directed to a wavemeter to monitor the wavelength and the other is sent to an optical spectrum analyzer to monitor the SMSR.

We first characterize thelectrical and optical performance of the laser chip with coarse tuning. Figure~\ref{fig:laser}(b) shows the light-current-voltage (LIV) curve corresponding output power as measured by the on-chip photodiode monitor (PD). The gain injection current was incrementally increased from $0$ to $80$ mA, and the PD response (blue trace) indicates a clear lasing threshold current near $32$ mA. As the injection current continues to ramp above the threshold, the integrated PD current steadily increases with some oscillations, indicating lasing mode hops with current tuning. This is a result of thermal heating and carrier injection in the active section that causes a change in the optical refractive index, producing a variation of in optical length of the cavity.

We next characterize the coarse tuning range of the laser chips. Optical spectra taken from two lasers with slightly different gain regions show the effectiveness of the Vernier ring configuration in maintaining a high SMSR over a wide wavelength range, shown in Fig.~\ref{fig:laser}(a). Only wavelengths resonant with the add-drop ring resonators in the Vernier mirror will overcome the cavity losses to lase; therefore, tuning the two rings in the Vernier mirror appropriately is crucial for single-mode operation. The rings were thermo-optically tuned independently to select a specific lasing mode across each III-V gain bandwidth. Stable single mode lasing across more than $17$~nm of continuous tuning range is demonstrated for each gain chip. The overlaid spectra are normalized to the peak power corresponding to each laser, leading to $>40$ dB SMSR and $>35$ dB SMSR for the blue and burgundy traces, respectively.

Lastly, the frequency noise was measured with a commercial OEwaves optical phase noise analyzer. Below $1$ MHz offset frequencies, noise is primarily due to environmental perturbations and the instrumentation, while above 1 MHz, the noise arises primarily from thermo-refractive effects. Frequency noise characterization further demonstrates a Schawlow-Townes fundamental linewidth of $5.85$ kHz, as shown in Fig.~\ref{fig:laser}(c). Further reduction in the laser linewidth can be achieved via self injection locking, using an external high quality factor ring resonator, resulting in a fundamental linewidth of $92.4$ Hz, as demonstrated by Zhang \textit{et al.} ~\cite{zhang2023photonic}.

\section{Mode-Hop-Free Tuning}

\begin{figure*}[t!] 
\centering
  \includegraphics[width=1\linewidth]{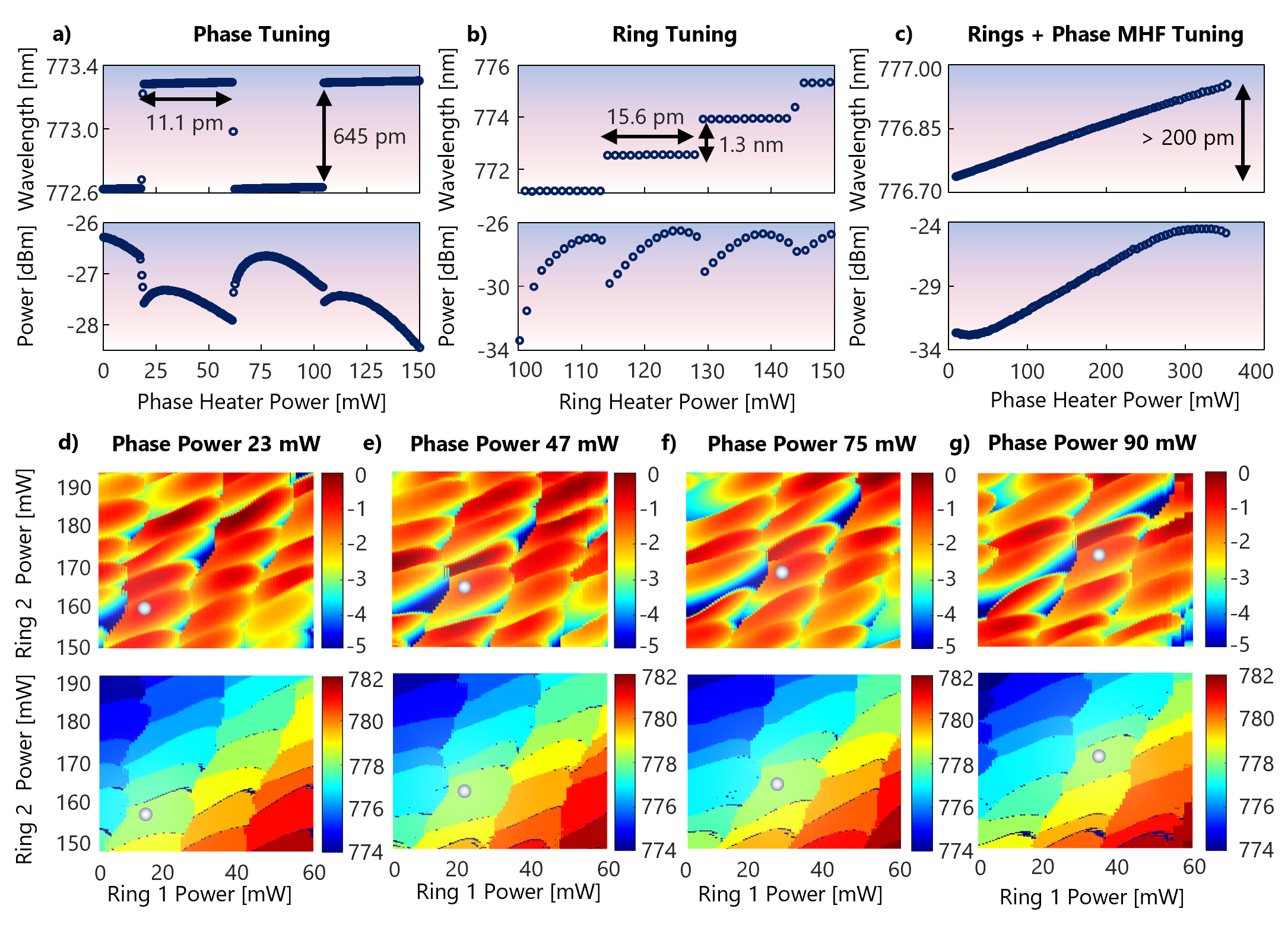}
  \caption{\footnotesize \label{fig:MHF} \textbf{Mode-hop-free tuning algorithm implementation.} (a) Isolated tuning of the phase heater demonstrating a narrow tuning range and mode hops at the cavity FSR. (b) Isolated tuning of one ring in the Vernier mirror, demonstrating a narrow tuning range within each step and mode hops at the Vernier mirror FSR. (c) Combined and synchronous tuning of the Vernier rings and phase heater to achieve $\geq$ 100 GHz (200 pm) of continuous tuning. Panels (d-g) show the 2D tuning maps for the output power (top row) and output wavelength (bottom row) versus the heater power on Ring 1 and Ring 2. As the phase heater power is incrementally increased, tracking of a local maximum of the output power, indicated by the highlight circle on each panel, enables the mode-hop-free tuning.}      
\end{figure*}

Continuous mode-hop-free tuning requires synchronous control of each tuning element in the laser cavity. To show this, we first demonstrate tuning by individually controlling each element. In Fig.~\ref{fig:MHF}(a), only the intracavity phase element is tuned while keeping the Vernier mirror heater power fixed. As the phase shifter power increases, a small tuning range up to $\sim$11~pm is observed until a mode hop of 645~pm occurs, likely between two Vernier peaks. Alternatively, Fig.~\ref{fig:MHF}(b) exhibits tuning of one ring of the Vernier mirror while keeping the phase-shifter element heater power fixed. In this case, a small tuning range up to $\sim$16~pm is observed until a large mode hop of 1.3~nm occurs, corresponding to the FSR of the ring.

Despite the limited continuous tuning range of each individual component, using an algorithmic approach to simultaneously tune the phase heater and both Vernier mirror ring heaters can enable tracking of a longitudinal mode---and thus continuous mode-hop-free tuning---over a much wider wavelength range. Here, we developed and implement a tuning algorithm that was originally demonstrated for a $1550$~nm InP/Si Vernier laser with 325 GHz MHF tuning \cite{pintus2023demonstration}. This method involves mapping the output power and wavelength of the laser as a function of the heater power applied to each Vernier ring and the phase shifter. These maps are then used to track a desired target mode over the 3D parameter space.

We first begin by generating a 2D map of the output power and wavelength of the laser as a function of applied power to Vernier mirror ring $\#1$ and ring $\#2$, as shown in Fig.~\ref{fig:MHF}(d-g). In Fig.~\ref{fig:MHF}(d), the phase heater power was fixed at 23 mW and the power was linearly tuned between 0-60 mW and 150-190 mW for ring $\#1$ and ring $\#2$, respectively . Each distinct region in the map reflects the overlap of a resonance from each ring with a particular longitudinal cavity resonance. Each region is characterized by a peak mode power at some specific ring heater power settings within the mode space. As the phase shifter power increases, the entire 2D map continuously shifts along the diagonal corresponding to the increase in applied heater power to ring $\#1$ and ring $\#2$. Thus, the key to continuous MHF tuning is to adjust the Vernier ring heaters and the phase shifter synchronously to follow the local maximum of the optical output power. To do this, we build a 3D map corresponding to the 2D maps shown as a function of the phase-shifter power (representative maps are shown in Fig.~\ref{fig:MHF}(d-g)). Cross-correlation of the peak mode powers in each 2D mode cloud map across successive phase power values then allow us to establish a linear relationship in applied power between the phase shifter heater and Vernier ring heaters for a specific mode.

Utilizing the established relationship between the ring heaters and the phase heater, we implemented the continuous tuning algorithm on several of the laser chips. We show in Fig.~\ref{fig:MHF}(c) representative results from one laser. By synchronously tuning the ring and phase heaters, we achieve 108~GHz (217~pm) of MHF tuning with 350~mW applied to the phase shifter. This range is a factor of 3 larger than previous hybrid-integrated devices and comparable to commercial tabletop ECDLs \cite{corato2023widely,winkler2024widely}. Limited to a maximum applied power of 1~W to the phase heater due to thermal load management, an estimated maximum mode-hop-free tuning range of 307~GHz (620~pm) is feasible for this design. MHF tuning of several THz should be possible through thermal isolation of the tuning elements using air trenches or through alternative tuning mechanisms such as electro-optic elements, as discussed in Section~5.

\section{Quantum Photonic Demonstrations}

\begin{figure*}[ht] 
\centering
     \includegraphics[width=1.0\linewidth]{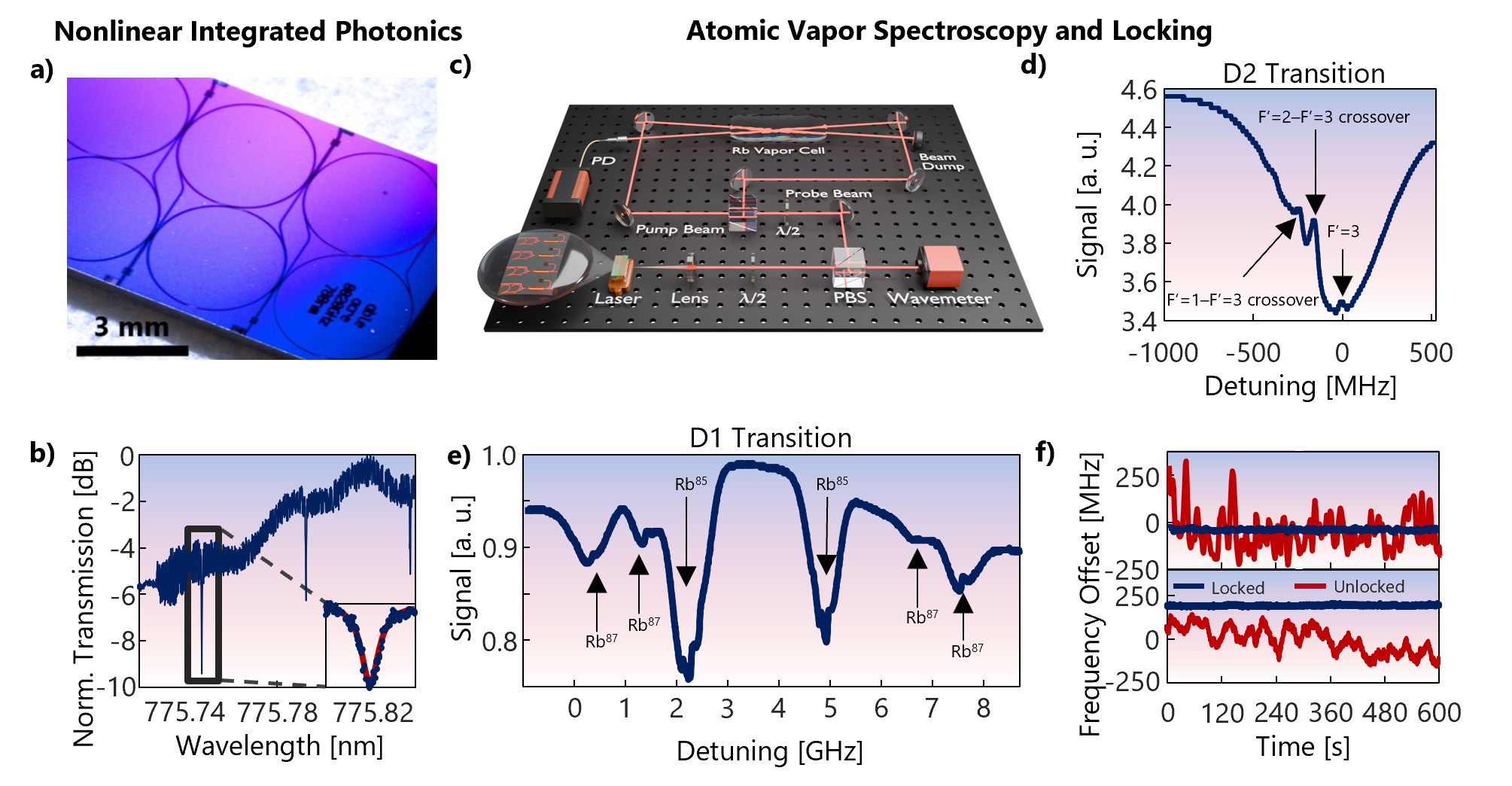} 
     \caption{\label{fig:Rb}\footnotesize  \textbf{780 laser applications in nonlinear photonics and atomic vapor spectroscopy and locking}. (a) SiN microring resonators with 20 GHz FSR used for testing the mode-hop-free tuning algorithm. In (b), the normalized linear transmission spectrum reveal several cavity modes, and a fit to the lineshape shown in the inset indicates an intrinsic cavity quality factor of 2.6 million. (c) A schematic of the experimental setup used to obtain and lock to a spectroscopy signal from a Rubidium vapor cell. (d) Representative signal from the $^{87}$Rb F = 2 D2 transition, and panel (e) shows the full spectroscopy signal from D1 for both $^{85}$Rb and $^{87}$Rb. (f) The wavelength of the lasing frequency over 10 minutes with and without locking to the sub-Doppler spectroscopy signal for both the D1 (top) and D2 (bottom) transitions.}
\end{figure*}

The development of MHF continuously tunable integrated lasers around 765 nm to 795 nm is motivated in part by numerous quantum-inspired applications. The most notable is the presence in this wavelength range of the two primary rubidium transitions---the D1 and D2 lines at 780 nm and 795 nm, respectively. Atomic systems based on Rb are being developed for several different quantum technologies, including quantum memories~\cite{gera2024hong}, entangled-photon pairs~\cite{craddock2024automated}, atomic clocks~\cite{martin2018compact}, and quantum sensors~\cite{cantat2020long,bai2021quantum,malia2020free}. Solid-state quantum systems with optical transitions in the near-IR, such as droplet epitaxy GaAs quantum dots~\cite{watanabe2000fabrication,schöll2019resonance}, quantum defects in atomically thin materials~\cite{azzam2021prospects}, and color centers in wide-gap materials like diamond and hexagonal boron nitride~\cite{lindner2018strongly,mohajerani2024narrowband,tran2018resonant}, also require MHF tunable lasers for optical excitation in the 775-795 nm wavelength range with narrow linewidths to measure and control the optical excitations resonantly. For nonlinear quantum photonics~\cite{moody2020chip}, 775-800 nm is a common optical pump wavelength for 1550-1600 nm entangled-photon pair generation via spontaneous parametric down conversion (SPDC) in $\chi^{\left(2\right)}$ materials, which has traditionally been performed with table-top optical pump lasers. Likewise, entangled-pair generation in the near-infrared via spontaneous four-wave mixing (SFWM) in integrated photonic micoresonators, such as SiN, is appealing for free-space and space-based quantum communications and entanglement distribution.

With these applications in mind, we initially focus on two demonstrations. The first---spectroscopy of a SiN microring resonator---is designed to highlight the MHF tuning capabilities for pumping integrated photonic devices for quantum light generation. The second---atomic spectroscopy and laser locking to the D1 and D2 lines---is an exemplary experiment to illustrate the ability for low-SWaP applications in atomic and molecular physics.

\subsection{780 nm Silicon Nitride Microresonator Characterization}

Our first demonstration is motivated by recent progress in the development of several integrated photonic material platforms for telecom-wavelength entangled-photon pair generation via SPDC, including lithium niobate \cite{zhao2020high} and indium gallium phosphide \cite{akin2024ingap,thiel2024wafer} and visible-wavelength SFWM \cite{garay2023fiber}. In these experiments, typically an ECDL laser is utilized as the pump to generated entangled pairs, which has been a limiting factor for transitioning quantum light source technologies from the lab into deployable, robust, and compact systems. Thus, an integrated 775 nm laser with MHF tuning, milliwatt output power, $>40$ dB SMSR, and sub-10 kHz linewidth will find immediate applications for quantum light generation; here, we demonstrate this possibility by pumping a SiN nonlinear resonator. For these measurements, a fraction of the pump power is routed to a wavemeter and a power meter before the chip, while the majority of the optical power is coupled into the SiN ring resonator. Lensed fibers are utilized to inject and collect transverse electric polarized light into and from the ring resonator. Light out of the chip is sent to a second power meter. The transmitted power is measured versus the laser wavelength.

In this case, the SiN resonator is designed for visible-wavelength pair generation via SFWM. The ring resonator cross section is 40~nm thick and 8~$\mu$m width with a radius of 1.49 mm, as shown in Fig.~\ref{fig:Rb}(a). In Fig.~\ref{fig:Rb}(b), we show a linear transmission spectrum acquired by continuously tuning the laser, which features multiple resonances in a single sweep that agrees with the 20 GHz FSR of the resonator. We fit a Lorentzian to the central resonance to determine loaded and intrinsic quality factors, which correspond to $2.5\times10^6$ (153 MHz linewidth) and $2.6\times10^6$ (144 MHz linewidth), respectively. 


\subsection{Rubidium Vapor Spectroscopy and Locking}

In our second demonstration, we use the integrated laser to perform Doppler-free spectroscopy on ground-state atomic transitions of rubidium. A stable narrow-linewidth laser is essential for clearly resolving the hyperfine structure in these transitions. The realization of saturated absorption spectroscopy also enables us to demonstrate a robust locking scheme in which the chip laser is stabilized to an absolute atomic frequency reference. Note that for these experiments, a variation of the laser cavity design with an etched-facet front mirror was used. The laser output is butt-coupled into the SiN waveguide from the front facet mirror.

The setup for the atomic spectroscopy experiments is shown by the schematic in Fig.~\ref{fig:Rb}(c). The laser light is first collimated, then passed through a half-wave plate (HWP) and a polarizing beamsplitter, which together serve as a tunable splitter to direct a fraction of the beam to a wavemeter, while the majority of the power is directed to a saturated absorption spectroscopy setup. Another HWP and polarizing beamsplitter are used to adjust the power appropriately between the pump and probe beams at approximately a 10:1 ratio. The nearly-counterpropagating pump and probe beams are directed to intersect at a shallow angle within a heated rubidium vapor cell to realize Doppler-free spectroscopy. Light from the probe beam is detected by a biased photodiode, the output signal from which is relayed to an oscilloscope and to locking electronics. When the laser detuning is scanned over the GHz-scale Doppler-broadened absorption dip of the D2 transition in $^{87}$Rb, saturated absorption shows clearly resolved transitions between different hyperfine states and crossover features (Fig.~\ref{fig:Rb}(d)). Scanning over a broader range at the 10 GHz scale near the D1 transition, achieved by using a 2-channel function generator to drive both Vernier rings simultaneously, the Doppler-broadened absorption dips from both stable isotopes of rubidium are clearly observed, each with its characteristic pattern of Doppler-free hyperfine and crossover features (Fig.~\ref{fig:Rb}(e)). These spectroscopic results demonstrate that the linewidth and tunability of the chip laser are on par with commercially available narrow-line laser systems used for precision experiments such as laser cooling.

This spectroscopic performance also enables the chip laser source to be stabilized to an absolute atomic frequency reference. To demonstrate this, we used a standard PID controller to provide feedback to the laser current from an error signal derived from the Doppler-free spectroscopy signal. This enabled drift-free stabilization of the laser frequency. The lock could be performed on both the D1 and D2 lines, and was stable over at least 10 minutes, as shown in Fig.~\ref{fig:Rb}(f).

\section{Conclusion}

In this work, we demonstrated an integrated III-V on silicon nitride laser platform operating from 765~nm to 795~nm with $\geq100$ GHz (200 pm) mode-hop free tuning, 6 kHz intrinsic linewidth, 40 dB side-mode suppression, and up to 10 mW output power. To the best of our knowledge, this represents the widest mode-hop-free tunability range for integrated lasers in this wavelength range, which is enabled by synchronously tuning the Vernier mirror and phase-shifter to track the local maximum power of the laser output.

In an exemplary demonstration of the laser platform capabilities, we first performed linear transmission spectroscopy of a silicon nitride microring resonator with a intrinsic quality factor $>2$ million, which required the large mode-hop-free tuning range and narrow linewidth to accurately map out several modes of the resonator. We next coupled the laser to a rubidium vapor cell in a Doppler-free configuration and performed atomic spectroscopy on the D1 and D2 atomic transitions, revealing clearly resolved hyperfine states and crossover features. By providing feedback to the laser current, the laser was absorption locked to both the D1 and D2 lines, providing a simple method for atomically referenced chip-scale lasers.

In the current study, the silicon nitride resonator and the atomic reference cell were external to the integrated laser platform. With recent advances in chip-scale optical isolators to prevent back reflections into the laser cavity~\cite{tian2021magnetic,white2023integrated,zhang2019monolithic}, additional components can be heterogeneously integrated onto the platform, including low-loss nonlinear waveguides and resonators for quantum light generation~\cite{baboux2023nonlinear}, quantum emitter chiplets for single-photon generation~\cite{larocque2024tunable,wan2020large}, and integrated atomic vapor cells for optical clocks and sensors~\cite{isichenko2023photonic,newman2019architecture}. The fully integrated platform makes the laser system robust to vibrations, shock, and temperature variations up to 110 \textdegree C ~\cite{zhang2023photonic}, with enhanced performance and stability expected in fully packaged devices. By optimizing the waveguide propagation loss, we anticipate that we can achieve sub-kHz linewidths, and injection locking with a $\sim10$ million quality factor resonator can further reduce the linewidth to sub-100 Hz~\cite{zhang2023photonic}. In the existing laser design, the mode-hop-free tuning range is limited primarily by thermal crosstalk, which can be improved by implementing lateral air trenches to thermally isolate each tunable component. The thermo-optic components can also be replaced with electro-optic elements, potentially based on the same semiconductor stack as the III-V gain region or through heterogeneous integration with lithium niobate, which could enable petahertz-per-second tuning speeds~\cite{snigirev2023ultrafast}.

\bigskip
\noindent\textbf{Funding.} A portion of this work was performed in the UCSB Nanofabrication Facility, an open access facility. This work was supported by the NSF Quantum Foundry through the Q-AMASE-i Program (Grant No. DMR-1906325), the NSF QuSeC-TAQS program (Grant No. 2326754), and the NSF CAREER Program (Grant No. 2045246). L.T. acknowledges support from the NSF Graduate Research Fellowship Program.


\bigskip

\noindent \textbf{Disclosures.} The authors declare no conflicts of interest.











\bigskip


\noindent\textbf{Data availability.} Data underlying the results presented in this paper are not publicly available at this time but may be obtained from the authors upon reasonable request.








\bibliography{Biblio}

\end{document}